\documentclass[aps,prl,superscriptaddress,notitlepage,twocolumn]{revtex4-1}

\usepackage{mathrsfs}
\usepackage{amsfonts}
\usepackage{amssymb}
\usepackage{amsmath}
\usepackage{bm}
\usepackage{graphicx}
\usepackage{sidecap}
\usepackage{color}
\usepackage[colorlinks=true,citecolor=blue,linkcolor=cyan]{hyperref}
\usepackage{subeqnarray}
\usepackage{verbatim}
\usepackage{ulem}
\usepackage{cases}
\usepackage{cancel}
\usepackage{framed}
\usepackage{tikz}

\newcommand{\ket}[1]{\vert #1 \rangle}

\begin{document}
\title{Protecting logical qubits with dynamical decoupling}
\author{Jia-Xiu Han}
\thanks{These authors contributed equally to this work.}
\affiliation{Beijing Academy of Quantum Information Sciences, Beijing 100193, China}
\affiliation{Hefei National Laboratory, Hefei 230088, China}
\author{Jiang Zhang}
\thanks{These authors contributed equally to this work.}
\affiliation{Beijing Academy of Quantum Information Sciences, Beijing 100193, China}
\author{Guang-Ming Xue}
\affiliation{Beijing Academy of Quantum Information Sciences, Beijing 100193, China}
\affiliation{Hefei National Laboratory, Hefei 230088, China}
\author{Haifeng Yu}
\email{hfyu@baqis.ac.cn}
\affiliation{Beijing Academy of Quantum Information Sciences, Beijing 100193, China}
\affiliation{Hefei National Laboratory, Hefei 230088, China}
\author{Guilu Long}
\email{gllong@mail.tsinghua.edu.cn}
\affiliation{Beijing Academy of Quantum Information Sciences, Beijing 100193, China}
\affiliation{State Key Laboratory of Low-Dimensional Quantum Physics and Department of Physics, Tsinghua University, Beijing 100084, China}
\affiliation{Frontier Science Center for Quantum Information, Beijing 100084, China}
\affiliation{Beijing National Research Center for Information Science and Technology, Beijing 100084, China}

\begin{abstract}
Demonstrating that logical qubits outperform their physical counterparts is a milestone for achieving reliable quantum computation.
Here, we propose to protect logical qubits with a novel dynamical decoupling scheme that implements iSWAP gates on nearest-neighbor physical qubits, and experimentally demonstrate the scheme on superconducting transmon qubits.
In our scheme, each logical qubit only requires two physical qubits. A universal set of quantum gates on the logical qubits can be achieved such that each logical gate comprises only one or two physical gates.
Our experiments reveal that the coherence time of a logical qubit is extended by up to 366\% when compared to the better-performing physical qubit.
Moreover, to the best of our knowledge, we demonstrate for the first time that multiple logical qubits outperform their physical counterparts in superconducting qubits.
We illustrate a set of universal gates through a logical Ramsey experiment and the creation of a logical Bell state. Given its scalable nature, our scheme holds promise as a component for future reliable quantum computation.

\end{abstract}

\pacs{}

\keywords{}

\maketitle

\textit{Introduction.---}Physical qubits are fundamental building blocks of quantum computers but are inherently susceptible to errors due to unavoidable interactions with their environment \cite{nielsen2010quantum,krantz2019quantum}. 
To perform reliable quantum computation, one needs to employ logical qubits which are defined using a set of physical qubits \cite{lidar2013quantum}.
It is a milestone to demonstrate that logical qubits outperform their associated physical counterparts, known as the ``break-even point'' \cite{devoret2013superconducting,ofek2016extending,sivak2023real,ni2023beating}.
One available approach to achieving this goal is to utilize quantum error correction (QEC) \cite{lidar2013quantum,devitt2013quantum,terhal2015quantum,hu2019quantum,campagne2020quantum,gertler2021protecting,krinner2022realizing,zhao2022realization,google2023suppressing}. 
Despite impressive achievements, connecting multiple logical qubits presents a significant challenge for these schemes.
To perform quantum algorithms with logical qubits before large-scale QEC are realized, it is still intriguing to investigate scalable error-mitigation or error-suppressing schemes for protecting logical qubits.

Dynamical decoupling (DD) (for reviews, see \cite{yang2011preserving,lidar2013quantum}) can manipulate interactions between a set of physical qubits and their environment by applying tailored control pulses to the qubits. 
One method is to completely eliminate these interactions, isolating the evolution of  qubits  \cite{viola1998dynamical,viola1999dynamical,viola2003robust,viola2005random,khodjasteh2005fault,yang2008universality,uys2009optimized,du2009preserving,khodjasteh2010arbitrarily,west2010near,de2010universal,rizzato2023extending}. 
Another method is to selectively eliminate non-collective interactions while preserving collective ones, in which the qubits interact with the environment in a uniform manner \cite{zanardi1999symmetrizing,viola2000dynamical,wu2005holonomic,zhang2019protection}. 
The latter introduces decoherence-free subspaces (DFS) and noiseless subsystems (NS), allowing logical qubits to be encoded, immune to collective errors. 
Previous pioneering work has demonstrated how to optimally generate collective interactions using DD sequences consisting of SWAP gates \cite{zhang2019protection}. 
In this scenario, all types of collective interactions are kept by SWAP operations, and thus rather complex encoding and decoding procedures are required to store and retrieve information with logical qubits. Logical operations on these logical qubits generally demand multi-body interactions, which are complicated to realize in practical. Moreover, SWAP gates cannot be directly implemented on various quantum computing platforms. As a result, an ideal DD scheme for protecting logical qubits is expected to utilize directly realizable quantum gates and generate DFSs in which logical qubits can be defined with simple encoding and decoding procedure as well as experimentally feasible logical gates.

In this work, we propose to generate $Z$-type collective interactions for physical qubits by periodically performing iSWAP gates. While the $Z$-type collective interactions are created, the $X$- and $Y$-type interactions are removed from the system evolution.
Compared with existing DD works in superconducting qubits \cite{bylander2011noise,gustavsson2012dynamical,guo2018dephasing,pokharel2018demonstration,sung2019non}, our scheme can be regarded as many-body DD \cite{qiu2021suppressing}.
The resulting $Z$-type collective interaction supports a DFS, enabling the encoding of logical qubits, each of which requires only two physical qubits. 
One feature of our scheme is scalability, and generating collective interactions for multiple qubits requires only iSWAP gates between nearest-neighbor qubits. 
Furthermore, universal gates for logical qubits can be realized using single-body and nearest-neighbor two-body Hamiltonians.
We experimentally demonstrate that the coherence time of a protected logical qubit is extended by up to 366\% compared to that of the better-performing physical qubit. 
For two logical qubits, the process fidelity is 194\% compared to that of the physical qubits at a duration of $\tau=4.8~\mu$s. To the best of our knowledge, this is the first time that multiple logical qubits have been shown to outperform their physical counterparts in superconducting qubits.
To showcase the universal control of logical qubits, we conduct a logical Ramsey experiment on one of the logical qubits and create a logical Bell state with logical operations.
Additionally, we demonstrate that our scheme can enhance the success probability in superdense coding as an application.

\begin{figure*}
	\centering
	\includegraphics{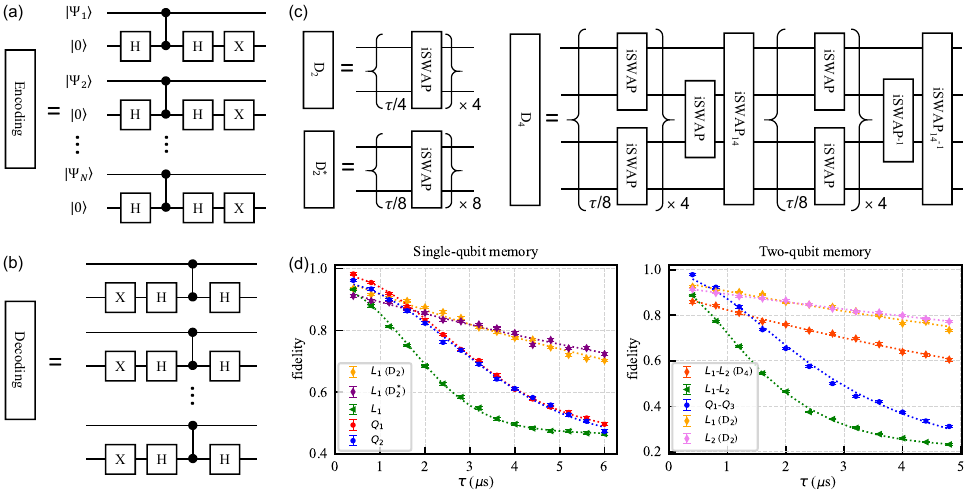}\\
	\caption{Quantum circuits and process fidelities for various procedures. (a) Encoding circuit for $N$-logical qubits. (b) Decoding circuit for $N$-logical qubits. (c) Quantum circuits of D$_2$, D$_2^*$, and D$_4$. (d) Left: Process fidelities of single-qubit identity operation over increased durations. Right: Process fidelities of two-qubit identity operation over increased durations. Details are shown in the inset. Dashed lines are fitting curves plotted using the fidelity functions \cite{SM}.}\label{fig:memory}
\end{figure*}

\textit{Protecting logical qubits with iSWAP gates.---}We begin by explaining how to encode a single logical qubit using a pair of physical qubits. When two qubits interact with their environment, the interaction Hamiltonian can be written as
\begin{equation}
	H_\text{I}=\sum_{\alpha=x,y,z}\sigma_{\alpha}^1\otimes E_{\alpha}^{1}+\sigma_{\alpha}^2\otimes E_{\alpha}^{2},
\end{equation}
where $\sigma_\alpha^k$ represents the Pauli-$\alpha$ operator acting on the $k$th qubit, and $E_\alpha^k$ is the related environmental Hamiltonian \cite{viola1999dynamical}. 
$H_\text{I}$ may cause dephasing and dissipation to the qubits, distorting the evolution of the physical qubits. 
The independent interaction $H_\text{I}$ can be transformed into a collective one using the DD procedure D$_2$ (shown in Fig.~\ref{fig:memory}(c)) which consists of four iSWAP gates applied in equal time interval. The obtained effective Hamiltonian $H^z_\text{I}=S_z\otimes \bar{E}_z$, with $S_z=(\sigma_z^1+\sigma_z^2)/2$ and $\bar{E}_z=E_z^1+E_z^2$, is a $Z$-type collective interaction \cite{SM}.

$H^z_\text{I}$ supports a DFS, denoted as $\text{DFS}_2=\text{Span}\{\ket{01},\ket{10}\}$, which serves as the encoding space for a logical qubit defined as $\ket{0}_L=\ket{01}$ and $\quad \ket{1}_L=\ket{10}$.
Since the two logical bases vanish under the action of $H^z_\text{I}$, they are dark states of $H^z_\text{I}$ and thus unaffected by it.
The encoding and decoding procedures of the logical qubits are explicitly shown in Fig.~\ref{fig:memory}(a) and (b).
Due to the fact that the encoding and decoding procedures of $\text{DFS}_2$ are concise, it is also used in other schemes, such as $g$-frame qubit \cite{qiu2021suppressing}, hardware-efficient architecture \cite{kapit2022small}, and dual-rail qubits \cite{teoh2023dual,levine2023demonstrating,chou2023demonstrating,kubica2023erasure}.
Two logical qubits can be encoded with four physical qubits in a four-dimensional DFS, denoted as DFS$_4=\text{Span}\{\ket{00}_L,\ket{01}_L,\ket{10}_L,\ket{11}_L\}$.
The DFS$_4$ is supported by a four-qubit $Z$-type collective interaction which can be generated with the procedure D$_4$ illustrated in Fig.~\ref{fig:memory}(c).
Our protective scheme is scalable in the sense that for $N$ logical qubits, we can also implement a DD procedure to generate a 2$N$-qubit $Z$-type collective interaction which supports a $2^N$-dimensional DFS \cite{SM}.

\begin{figure}[t]
	\centering
	\includegraphics[width=0.45\textwidth]{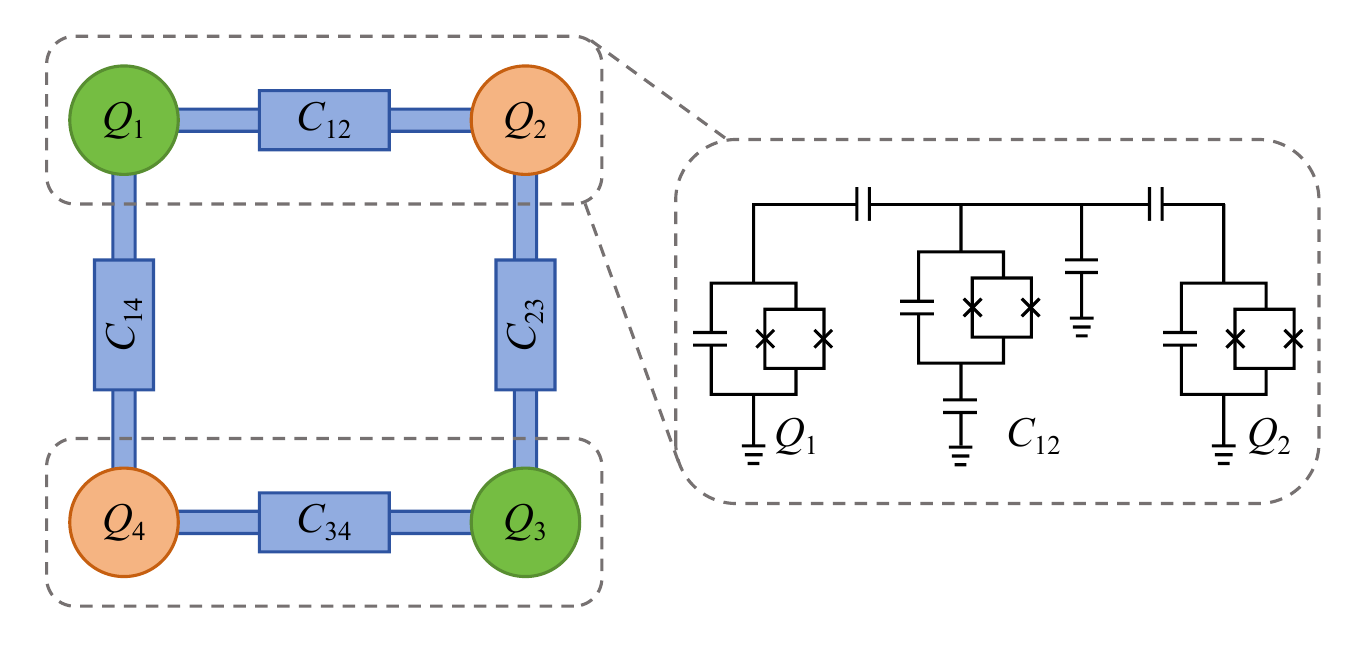}\\
	\caption{Arrangement of physical and logical qubits. Left: the physical qubits are denoted by green and orange circles. Interactions ($C_{12}$ to $C_{14}$) between physical qubits are labeled with blue rectangles. Physical qubits $Q_1$ and $Q_2$ are used to define the logical qubit $L_1$, while $Q_3$ and $Q_4$ constitute the logical qubit $L_2$. Right: the schematic circuit of the device including two transmon qubits and one tunable coupler.}\label{fig:device}
\end{figure}

\textit{Experimental demonstration of protecting logical qubits.---}In our experiments, we utilize a superconducting device consisting of four frequency-tunable transmon qubits.
Fig.~\ref{fig:device} shows the configuration of the device: the four physical qubits $Q_1$ to $Q_4$ are arranged in a circle with only nearest-neighbor coupling, and adjacent qubits are connected through tunable couplers \cite{yan2018tunable}, labeled as $C_{12}$ to $C_{14}$.
Each qubit connects to a dedicated driving line for individual XY control. 
The coupling strength between neighboring qubits can be adjusted from negative to positive by tuning the couplers' resonance frequencies.
This adjustability ensures the high-contrast on-off switching of inter-qubit coupling, which is essential for implementing entangling gates such as iSWAP and controlled-Z (CZ) gates \cite{foxen2020demonstrating}.
Additionally, we employ impedance-matched Josephson parameter amplifiers (JPAs) to achieve high-fidelity single-shot readout of the qubits \cite{roy2015broadband}.

We assess the memory times of both logical and physical qubits by employing quantum process tomography \cite{nielsen2010quantum} to quantify the fidelities of identity operation across various durations (for details, see \cite{SM}).
We begin by evaluating the performance of the logical qubit $L_1$ and its physical counterparts.
From Fig.~\ref{fig:memory}(d), one can see that without any protection mechanisms, the process fidelities of physical qubits $Q_1$ and $Q_2$ noticeably decrease with increasing duration $\tau$. The fidelity for the physical qubit $Q_i$ can be written as \cite{pedersen2007fidelity,magesan2011gate,SM}
\begin{equation}
\label{eq:physical_fidelity}
F_{Q_i}=\frac{1}{4}\left(2e^{-(\tau/\mathrm{T}_{2,i})^2}+e^{-\tau/\mathrm{T}_{1,i}}+1 \right),
\end{equation}
where $\mathrm{T}_{2,i}$ and $\mathrm{T}_{1,i}$ represent the coherence and energy relaxation times of physical qubit $Q_i$, respectively.

Without protection, $L_1$ experiences a more rapid fidelity degradation than a single physical qubit since two physical qubits introduce more errors, with the unprotected logical coherence time $\mathrm{T}_2^{\text{un}}=\mathrm{T}_{2,1}\mathrm{T}_{2,2}/(\mathrm{T}_{2,1}+\mathrm{T}_{2,2})$ and the energy relaxation time T$_{1,1}$. The fidelity of unprotected $L_1$ is affected by T$_1$ error only from $Q_1$ due to the tailored decoding procedure \cite{SM}.

Remarkably, when $L_1$ is protected by $\mathrm{D}_2$, the process fidelity of $L_1$ decreases much more slowly than that of unprotected $L_1$ as well as that of any physical qubits.
For instance, at $\tau=6.0~\mu$s, the process fidelity of $L_1$ remains 72.31\%, significantly higher than the 49.59\% observed for the better-performing physical qubit $Q_1$.
Compared with the Gaussian decay observed in the fidelity of physical qubits, the protected $L_1$ exhibits an exponential decay in process fidelity
\begin{equation}
\label{eq:logical_1Q_DD}
F_{L_1}=\frac{1}{4}\left(2e^{-\tau/\mathrm{T}_2^\text{p}}+e^{-\tau/\mathrm{T}_{1,1}}+1 \right),
\end{equation}
since the environment around $L_1$ is modified by D$_2$.
A prolonged logical coherence time $\mathrm{T}_2^\text{p} = 9.2~\mu$s is found from $F_{L_1}$, far exceeding the coherence times of physical qubits $\mathrm{T}_{2,1}=3.8~\mu$s and $\mathrm{T}_{2,2}=4.2~\mu$s. 

We also implement a procedure $\mathrm{D}_2^*$ which uses eight iSWAPs to protect $L_1$, shown in Fig.~\ref{fig:memory}(c). The application of D$_2^*$ increases the logical coherence time to $\mathrm{T}_2^\text{p}=12.5~\mu$s. 
Hence, our results convincingly verify that the memory of the protected logical qubit surpasses that of the better-performing physical qubit. We expect that longer logical coherence time can be obtained with more iSWAP gates applied in a protecting procedure.

\begin{figure*}
	\centering
	\includegraphics{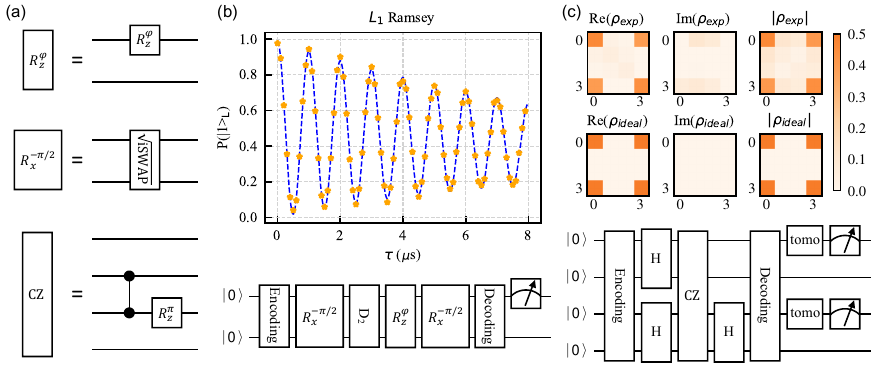}\\
	\caption{Universal logical operations. (a) The set of universal logical gates. (b) A Ramsey experiment is conducted on logical qubit $L_1$. The orange stars represent experimental data points, while the blue dashed line indicates the simulation results. (c) Preparation of the logical Bell state $B_1=(\ket{00}_L+\ket{11}_L)/\sqrt{2}$ from initial state $\ket{00}_L$ using logical Hadamard and CZ gates.
	The obtained density matrix elements are shown in the top, where 0 to 3 represent the bases $\ket{00}$, $\ket{01}$, $\ket{10}$, and $\ket{11}$, respectively.}\label{fig:operation}
\end{figure*}

We further evaluate the performance of two logical qubits protected by $\mathrm{D}_4$ shown in Fig.~\ref{fig:memory}(c). 
To achieve precise quantum operations, we reconfigure the operating frequencies of the four physical qubits and recalibrate the associated gates.
Illustrated in Fig.~\ref{fig:memory}(d), we replicate the single-logical-qubit experiments with procedure D$_2$ on both $L_1$ and $L_2$, resulting in a maximum 366\% increase in the coherence time of logical qubit $L_2$ ($15.0~\mu$s), compared to the coherence time of the better-performing physical qubit $Q_4$ ($4.1~\mu$s).
Moreover, the two-logical-qubit process, protected by $\mathrm{D}_4$, exhibits significant improvement over the two-physical-qubit process, as shown in Fig.~\ref{fig:memory}(d). 
When $\tau=4.8~\mu$s, the fidelity of the former remains 60.51\%, significantly surpassing that  of the latter (31.19\%).

In our experiments, we observe a trade-off between mitigating environmental errors and introducing control errors through the protection procedure.
Here, we chose to employ unoptimized quantum gates, because we are more concerned about the declining trend of process fidelity.
It is reasonable to anticipate that utilizing quantum gates with higher fidelity could shift the onset of this trade-off to an earlier time, potentially enhancing the fidelity of logical qubits. 

\textit{Universal operations for logical qubits.---}Our scheme enables universal control of logical qubits using single-body and nearest-neighbor two-body Hamiltonians (see Fig.~\ref{fig:operation}(a)). 
The logical Pauli-$z$ operator for $L_1$ ($\sigma_z^{L_1}$) is just the physical $\sigma_z^1$ operator. 
Then, for logical qubit $L_1$, a logical rotation $R_z^\varphi$ about the $z$-axis by an angle $\varphi$ is equivalent to the same rotation of physical qubit $Q_1$.
The logical Pauli-$x$ operator for $L_1$ ($\sigma_x^{L_1}$) is represented by the XY-interaction $\sigma_x^1\sigma_x^2+\sigma_y^1\sigma_y^2$.
It follows that the $\sqrt{\mathrm{iSWAP}}$ gate acting on $Q_1$ and $Q_2$ is the logical rotation $R_x^{-\pi/2}$ about the $x$-axis by an angle $-\pi/2$ for $L_1$.
With these two logical rotations available, we can perform an arbitrary operation, such as the logical Hadamard gate (H), on a logical qubit.
Additionally, the $\sigma_z^{L_1}\sigma_z^{L_2}$ interaction between $L_1$ and $L_2$ corresponds to the physical $\sigma_z^{2}\sigma_z^{3}$, which is a nearest-neighbor interaction between $Q_2$ and $Q_3$. 
Hence, a logical CZ gate can be achieved by implementing a CZ gate between $Q_2$ and $Q_3$ followed by a $R_z^\pi$ gate on $Q_3$. 
As a result, we can establish a complete set of universal logical gates: \{$R_z^\varphi$, $R_x^{-\pi/2}$, CZ\}.

To demonstrate single-qubit logical operations, we conduct a logical Ramsey experiment on $L_1$, incorporating the logical operations depicted in Fig.~\ref{fig:operation}(b). 
The experimental data closely aligns with the simulation results, which are calculated based on the energy relaxation time $\mathrm{T}_{1,1}$ and the coherence time $\mathrm{T}_2^{\text{p}}$ as input parameters.

The universal logical control is finally illustrated by preparing a logical Bell state $B_1=(\ket{00}_L+\ket{11}_L)/\sqrt{2}$ on $L_1$ and $L_2$, shown in Fig.~\ref{fig:operation}(c).
We utilize quantum state tomography to evaluate the state's fidelity, which is determined to be 92.30\%.
Our analysis indicates that about 28\% of the errors originate from the state tomography and readout process. 
The remaining errors predominantly arise from operations involving physical two-qubit gates ($\sqrt{\text{iSWAP}}$ and CZ). 
We stress that the main focus here is to show that universal control is available in our scheme. Hence, we choose a rather tedious circuit which comprises 7 layers and 11 two-qubit gates to create $B_1$. In practice, a more concise method is to prepare a Bell state for two physical qubits and then apply the encoding procedure. The latter method uses much fewer two-qubit gates and thus will obtain a higher state fidelity for $B_1$.

\begin{figure}
	\centering
	\includegraphics{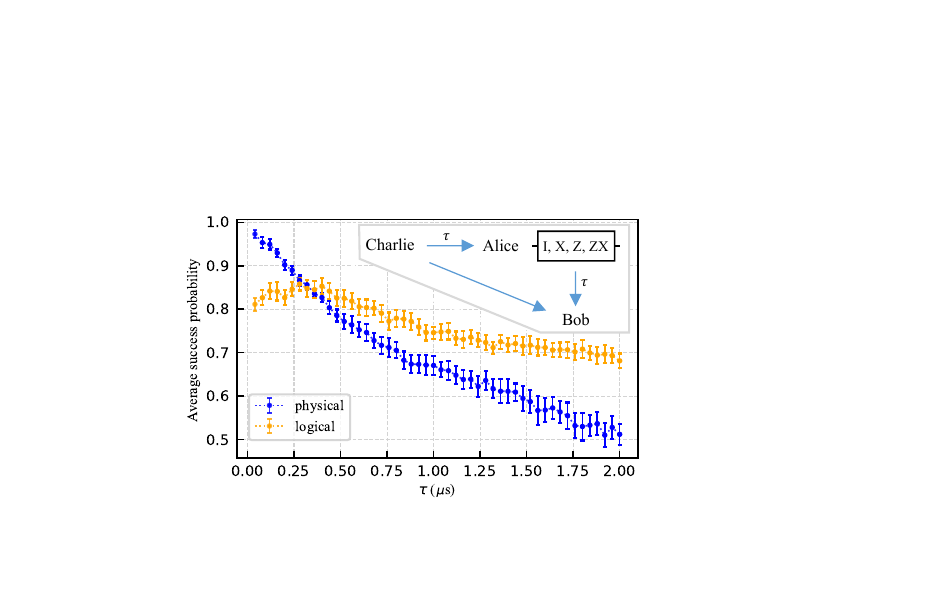}\\
	\caption{Superdense coding. The average success probabilities for logical superdense coding (illustrated in orange) and physical superdense coding (illustrated in blue).}\label{fig:superdense}
\end{figure}

\textit{Application to superdense coding.---}As an application, we demonstrate the utility of logical qubits in enhancing the performance of superdense coding \cite{Bennett1992Communication}. The process is illustrated in the inset of Fig.~\ref{fig:superdense}.

Initially, a trusted third party Charlie prepares the Bell state $B_1$. Then, Charlie sends one qubit to Alice and the other to Bob, assuming that this transmission takes time $\tau$. Upon receiving her qubit, Alice applies one of the four possible operations ($\{\text{I,X,Z,XZ}\}$) to it and then forwards it to Bob, which process takes the other $\tau$.
Bob then performs a Bell measurement, yielding four possible outcomes according to Alice's operations. This allows Bob to extract the information transmitted by Alice after a total duration of $2\tau$. We vary the duration $\tau$ to observe its impact on Bob's average success probability.

The experiments are conducted with both physical and logical qubits. For short $\tau$, the physical qubits outperform the logical ones because more control errors are introduced to the logical qubits. However, in a relatively long $\tau$, the logical qubits surpass the physical ones.

\textit{Discussion and Conclusion.---}Before proceeding further, we would like to provide some remarks on the differences between our scheme and existing works. First, widely used single-body DD schemes, such as XY-$N$ ($N=4,8$ and others) sequences, are unsuitable for protecting logical qubits since they can flip the logical states out of the DFS. However, combining the well-developed high-order DD \cite{khodjasteh2010arbitrarily} and optimal control \cite{zhang2023coupling} with our scheme may enhance its performance.
Secondly, the code in our scheme shares similarities with some recent works \cite{kapit2022small,teoh2023dual,levine2023demonstrating,chou2023demonstrating,kubica2023erasure}. 
The key feature of our scheme is that quantum states can be transferred from a logical qubit to one of its physical counterparts and vice versa, making our scheme compatible with other logical qubit protection schemes, such as QEC.

Our scheme can protect logical qubits from T$_2$ errors, but cannot eliminate T$_1$ errors.
In the next step, we may use the second physical qubit in a logical qubit as a herald qubit and apply feedback control to compensate it. 
Then, one may consider developing dynamically protected logical gates for the logical qubits to perform practical quantum algorithms.

In conclusion, we have proposed and verified protection of logical qubits using DD.
We have explicitly demonstrated that the lifetimes of both single- and two-logical qubits are prolonged compared to those of their physical counterparts.
Also, we conducted a logical Ramsey experiment and successfully created a logical Bell state to illustrate the universal control of the logical qubits.
Furthermore, we have shown that our scheme can enhance the success probability in superdense coding.

\begin{acknowledgments}
	We acknowledge fruitful discussions with Feihao Zhang.
	J.H. acknowledges support by the National Natural Science Foundation of China (No. 12004042).
	J.Z. acknowledges support by the National Natural Science Foundation of China (No. 12004206).
	H.Y. acknowledges support by the Innovation Program for Quantum Science and Technology (No. 2021ZD0301800) and National Natural Science Foundation of China (No. 92365206).
	G.L. acknowledges support by the National Natural Science Foundation of China (No. 62131002), Beijing Advanced Innovation Center for Future Chip (ICFC), and Tsinghua University Initiative Scientific Research Program.
\end{acknowledgments}

\bibliography{references}
\bibliographystyle{apsrev4-1}

\end{document}